\documentclass{elsart5p}

\usepackage{graphics}
\usepackage{graphicx}
\usepackage{subfig}

\usepackage{amssymb}

\usepackage{lineno}

\begin{document}

\begin{frontmatter}


\title{Progress Towards A Permanent Octupole Magnetic Ultra-Cold Neutron Trap for Lifetime Measurements}

\author[ILL,TUM]{K.~K.~H.~Leung\corauthref{cor}},
\corauth[cor]{Corresponding author.} 
\ead{leung@ill.eu}
\author[ILL,TUM]{O.~Zimmer}.

\address[ILL]{Institut Laue Langevin, 6, rue Jules Horowitz, 38042 Grenoble, France.}
\address[TUM]{Physik Department E18, Technische Universit\"{a}t M\"{u}nchen, James-Franck-Str.,85748 Garching, Germany}


\begin{abstract}

Current knowledge of the neutron $\beta$-decay lifetime has come under scrutiny as of late due to large disagreements between recent precise measurements. Measurements using magnetically trapped Ultra-Cold Neutrons (UCNs) offer the possibility of storage without spurious losses which can provide a reliable value for the neutron lifetime. The progress towards realizing a neutron lifetime measurement using a Ioffe-type trap made with a Halbach-type permanent octupole magnet is presented here. The experimental procedure extracts into vacuum UCNs produced from a superfluid helium converter and detects the neutron decays via \emph{in-situ} detection of produced protons.

\end{abstract}

\end{frontmatter}

\section{Introduction}
\label{sec:intro}

The lifetime of the free neutron ($\tau _{\mathrm{n}}$) is of importance in
particle physics and in cosmology. When combined with the ratio of the
number of baryons per photon ($\eta _{10}$), it leads to predictions of
experimentally testable quantities, such as the primordial abundances of $^{4}$He and deuterium \cite{Coc2007a}. In particle physics, the combination
of $\tau _{\mathrm{n}}$ with the neutron $\beta $-asymmetry parameter ($A$)
and the Fermi weak coupling constant ($G_{\mathrm{F}}$) from $\mu $-decay
provides a value for the quark mixing matrix element ($V_{\mathrm{ud}}$) in
the Cabibbo-Kobayashi-Maskawa (CKM) matrix. Unitarity tests using the first
row of the CKM-matrix offer a sensitive search for physics beyond the
Standard Model \cite{Ramsey-Musolf2008a}. Currently the most precise value
of $V_{\mathrm{ud}}$ comes from superallowed $\beta $-decay data \cite
{Hardy2005}. However, a value of $V_{\mathrm{ud}}$ from neutron decay data
contains an order of magnitude smaller theoretical corrections than from
nuclear structure dependent inner radiative processes. 

The accuracy of the $V_{\mathrm{ud}}$ determination from neutron decay would
be dominated by the experimental uncertainty in $A$ (which is being improved
in present measurements \cite{Abele2008a}) if being combined with the world
average value of $\tau _{\mathrm{n}}$ accepted prior to 2005, $885.7\pm 0.8\,
\mathrm{s}$ with $\chi _{\nu }^{2}=0.76$ \cite{Severijns2006a}. However,
this value has become rather suspect in the past few years \cite
{Yao2006a,Paul2008a}. It is dominated by a single measurement which used
material storage of ultra-cold neutrons (UCNs) \cite{Arzumanov2000a}. A
recent measurement in 2005 produced a result which was 7.2$\,\mathrm{s}$
lower and disagreed by 6.5 $\mathrm{\sigma }$ with the previous world
average \cite{Serebrov2005}. Since both these measurements were performed
using bottled UCNs, where interactions of UCNs with material walls contain
spurious losses due to either absorption or up-scattering, it is desirable
to employ an alternative technique which avoids potential systematics
associated with such losses.

Magnetic fields can be used for confinement of neutrons thereby avoiding the
need for material interactions \cite{Paul1989a,Vladimirskii2012a}. Since
this technique contains different systematic effects than what is present in
the two previous conflicting results, it should be able to resolve the
current problem. Furthermore, we argue that the amount of systematic
corrections and reliability in the technique presented is sufficiently small
such that a lifetime value with precision $<0.5\,$s can be achieved. This accuracy should be aimed for to match the precision in primordial abundance calculations \cite{Lopez1999a} and future measurements of $A$ \cite{Abele2008a}.

\section{Octupole Magnetic Trap Setup}
\label{sec:trap_setup}

Neutrons, with magnetic moment $\vec{\mu}_{\mathrm{n}}$ with magnitude $\mu
_{\mathrm{n}}=60\,\mathrm{neV/T}$, in a magnetic field $\vec{B}$, experience
a force $\vec{F}=\nabla (\vec{\mu}_{\mathrm{n}}\cdot \vec{B})$. Neutrons
with magnetic moment anti-parallel to the magnetic field, the so called 
\emph{low-field seekers}, will be forbidden to enter regions where the total
energy of the neutron $E\,$$>$$\,\mu _{\mathrm{n}}B$. This allows one to
confine neutrons with energy in the UCN-range with magnetic fields on the
order of several tesla. Three-dimensional magnetic storage using a
Ioffe-type trap built from superconductors was demonstrated in Ref. \cite{Huffman2000a}.

With the advent of modern magnetic materials, field strengths beyond one tesla can now be achieved with permanent magnets, allowing construction of traps with large volumes at a moderate cost. This was demonstrated recently by a small prototype magneto-gravitational UCN trap made with permanent magnets providing a nominal trapping field of 1$\, \textrm{T }$\cite{Ezhov2005}. We plan on constructing an Ioffe-type trap with a combination of permanent magnets and superconductors.

\begin{figure}
\begin{center}
\includegraphics[width=0.35\textwidth]{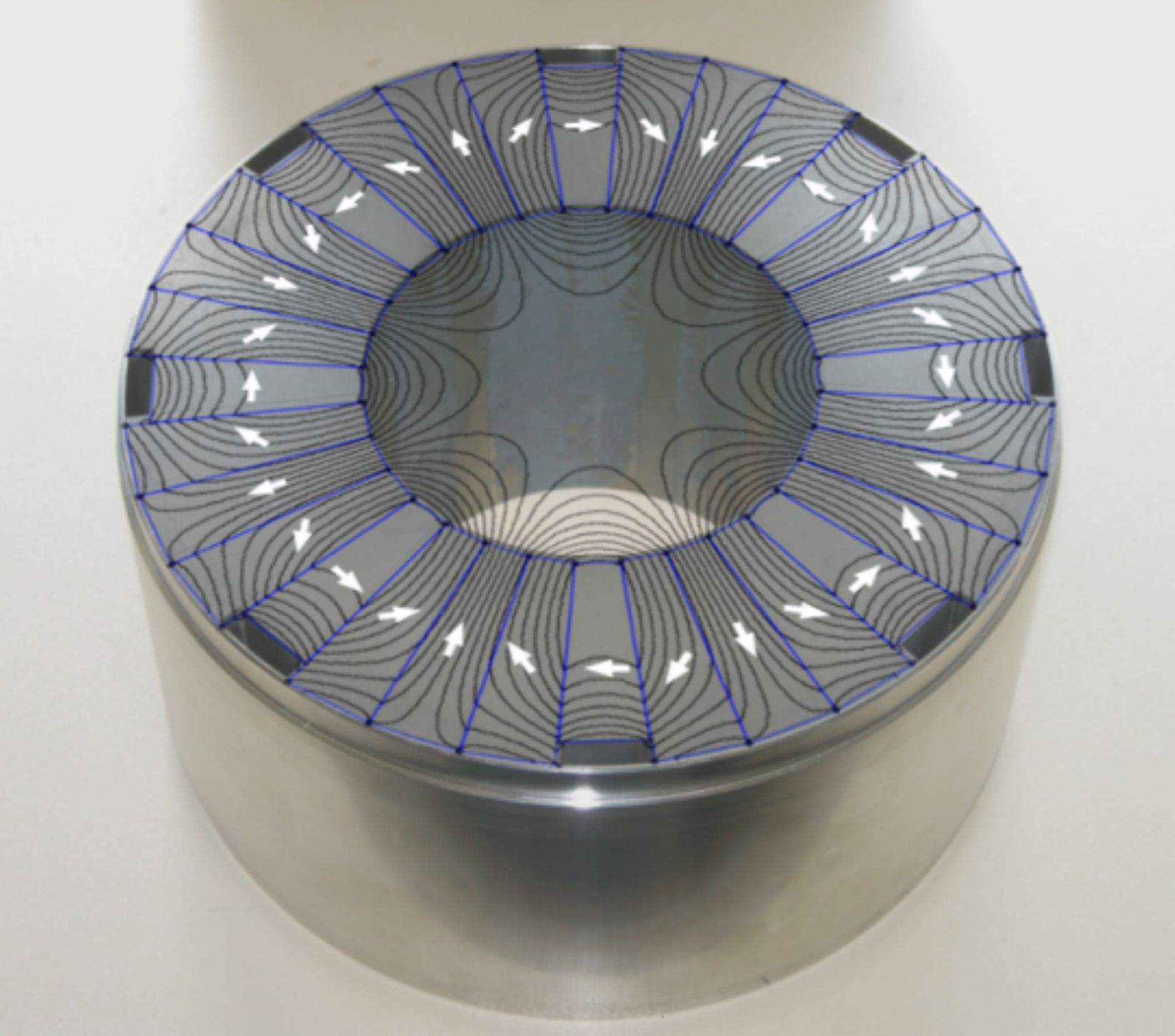}
\end{center}
\caption{One module of the Halbach-type octupole permanent magnet assembly (with its stainless steel enclosure). Magnetic field lines from finite element calculations are shown super-imposed. The arrows indicate the magnetization axis of the different segments.}
\label{fig:magnetFEMM}
\end{figure}

For radial UCN confinement a $1.2\,\textrm{m}$ long Halbach-type octupole \cite{Halbach1980} made from magnetized rare-earth material (NdFeB) is used (see Fig.~\ref{fig:magnetFEMM}). The magnet assembly has an inner bore diameter of $9.4\,\textrm{cm}$ and an outer diameter of $19\,\textrm{cm}$ and is constructed with $12$ identical units each $10\,\textrm{cm}$ long. The minimum field strength at the inner surface of the octupole slightly exceeds $1.3\,\textrm{T}$ at room temperature\footnote{This value will increase to almost 1.5$\,\textrm{T}$ by cooling the material to 110-140$\,\textrm{K}$. For subsequent estimates a conservative value of 1.4$\ \textrm{T}$ is used.}.

The trap will be closed off by two superconducting end solenoids with a large (30\,\textrm{cm}) diameter to reduce the partial cancellation of the transverse component of this field with the field of the octupole. There will also be a bias field solenoid (nominally $0.3\,$T) used to both prevent the neutrons from depolarizing and to help extraction of protons. Since the bias field is orthogonal to the octupole field, the effective trapping depth reduces to $1.1\,\mathrm{T}$. The magnetic field configuration is similar to that in Ref. \cite{Huffman2000a}. Note that, however, a radial octupole field is used instead of a quadrupole, resulting in an effective phase space trapping volume equivalent to that of a quadrupole trap of the same size with $1.8$ times higher trapping depth.

\begin{figure*}
\begin{center}
\includegraphics[width=1\textwidth]{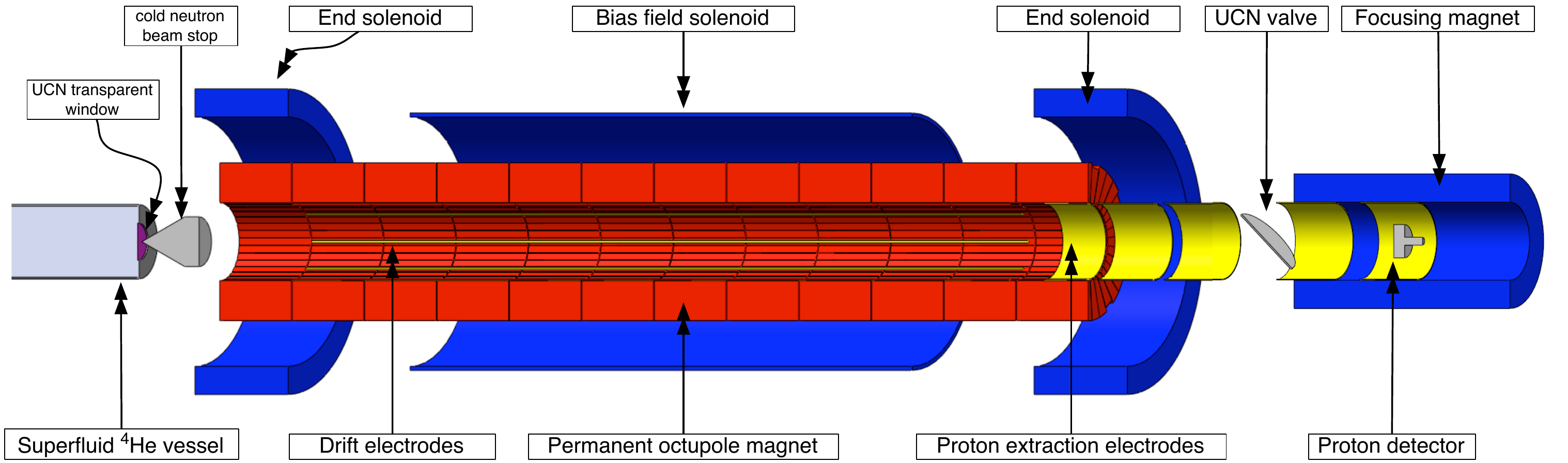}
\end{center}
\caption{Diagram of the lifetime experimental setup. The full system involves the permanent octupole magnet, superconducting solenoids and proton electrode system. The beam stop slides along with the superfluid $^4$He vessel.}
\label{fig:trap3D}
\end{figure*}

\section{Neutron Lifetime Measurement}
\label{sec:lifetime_measure}

\begin{figure}
\begin{center}
\includegraphics[width=0.47\textwidth]{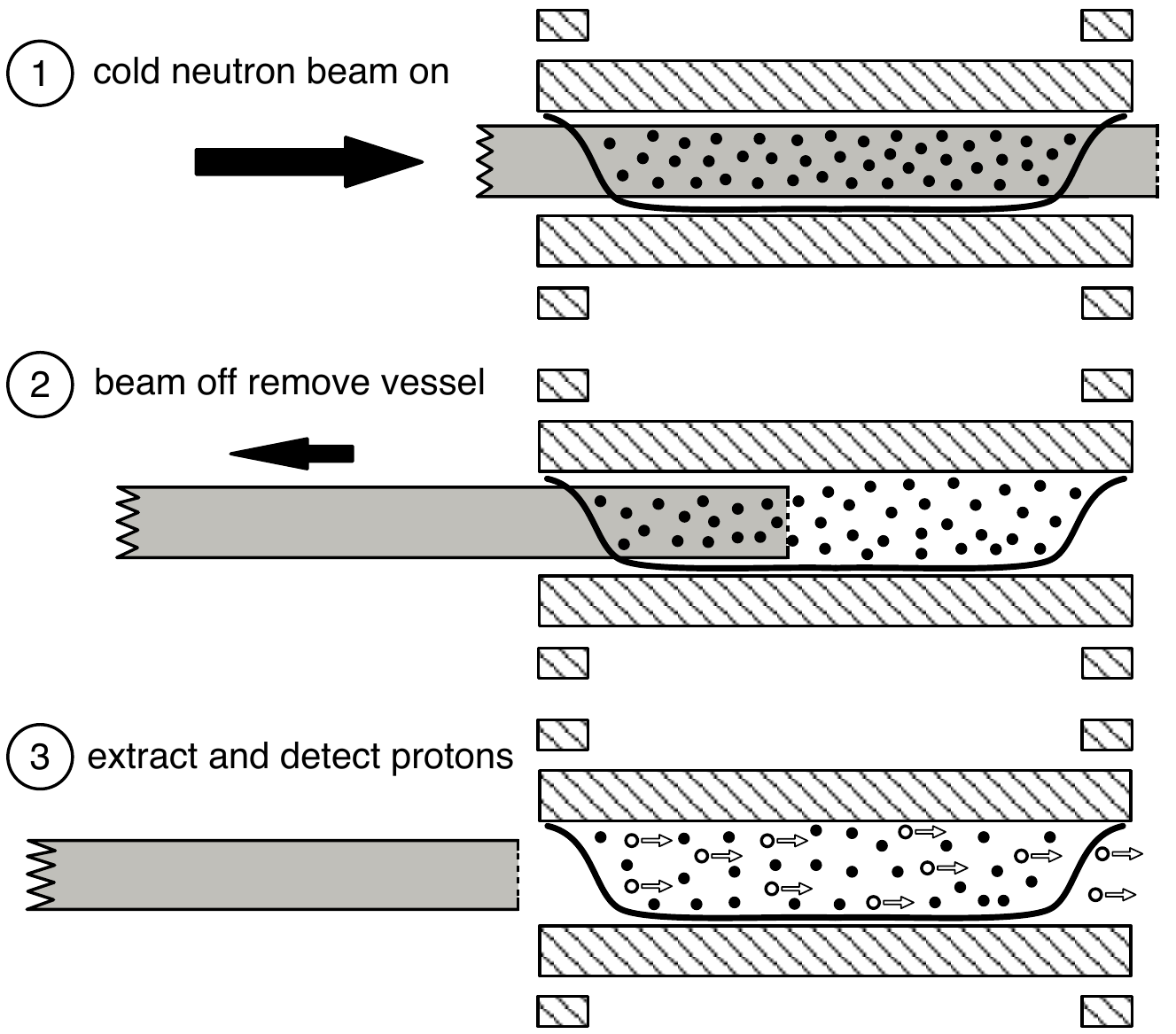}
\end{center}
\caption{Schematic describing the procedure of extracting the magnetically
trapped UCNs produced via. superthermal down-conversion into vacuum. Only
the octupole magnet, end solenoids and superfluid $^4$He vessel are shown.}
\label{fig:procedure}
\end{figure}

For UCN production we plan to take advantage of the high UCN density offered by a converter of isotopically purified superfluid $^{4}$He at a temperature of $\sim 0.5 \,$K in combination with an intense beam of cold neutrons \cite{Golub1975a}. The UCNs are produced inside the trapping volume by placing a cylindrically shaped converter vessel inside the bore of the octupole magnet (see Fig. \ref{fig:procedure}). After sufficient accumulation, the vessel, with an extraction window, will be slid out of the magnetic trap\footnote{In the actual experiment the magnetic trap is moved instead for practical reasons.} leaving behind the gas of UCNs. The decay of the UCNs in vacuum will be detected \emph{in-situ} by extraction of produced protons (described in Sec. \ref{sec:detection_protons} ). The silicon drift detector currently employed in the $a$SPECT experiment \cite{Simson2007a} provides a sharp proton signal well separated from electronic noise.

A suitable choice for the extraction window might be aluminum due to its low Fermi potential and low absorption. The speed of vessel retraction must not be so fast as to increase energy of the neutrons due to reflections off the moving potentials, but also be sufficiently fast to avoid repeated passages of the neutrons through the foil where absorption can occur. The optimum conditions still need to be investigated.

Although significant losses are expected in the extraction process, lifetime measurements outside the superfluid is crucial as it removes the effects of spurious neutron losses due to interactions with traces of $^{3}$He or with excitations in the $^{4}$He superfluid. For instance in a scheme where storage was measured inside superfluid $^4$He purified with superleaks, the uncertainty in the systematic correction required due to interactions with trace $^3$He (reduced to a level of $10^{-12}$) was estimated to be $\sim$70$\,$s \cite{Yang2006}. Sophisticated methods are required to reduce the trace concentration by three orders of magnitude before the systematic contribution becomes sufficiently small.

Monitoring neutron losses such as from depolarization or escape of above threshold neutrons can be achieved in the current setup by placing a neutron detector in between the end solenoid and the proton detector. This requires the inner surface of the trapping volume to be coated with an efficient UCN reflector. Compared with having a neutron absorber, the presence of a reflector increases the cleaning time of neutrons in quasi-stable orbits, since wall collisions which would have otherwise absorbed these neutrons, instead could reflect them back into other quasi-stable orbits. Therefore monitoring of losses will only be done in searches for systematic effects and not in every lifetime measurement run.

\emph{In-situ} measurements allow a lifetime value to be extracted from each individual filling. This compared to traditional \emph{fill-and-empty} techniques has the advantages that: (1) the same statistical uncertainty can be reached in a shorter time period, (2) analysis of the storage time at different time intervals allows for the search of systematic errors in data directly, (3) pile-up losses are avoided in the counting procedure, and (4) there is no need to assume that the initial number of neutrons in the trap is constant.

\section{Detection of Protons}
\label{sec:detection_protons}

The protons from neutron $\beta $-decay have a maximum kinetic energy of $\sim 0.75\,$keV. In order to extract them from the magnetic trap for detection, a series of electrodes (see Fig.~\ref{fig:trap3D}) are used to accelerate them over the magnetic mirror at the end solenoid and at an additional coil used for adiabatically focussing the protons onto a small detector. In the extreme case where the proton starts in the weakest field in the trap ($0.3$ T) and has all its kinetic energy in the motion transverse to the field axis (i.e. $E_{\bot }=0.75\,\mathrm{keV}$), the required voltage to overcome a magnetic mirror of $1.5\,$T is $\sim 4$ kV.

\begin{figure}[tbp]
\begin{center}
\includegraphics[width=0.45\textwidth]{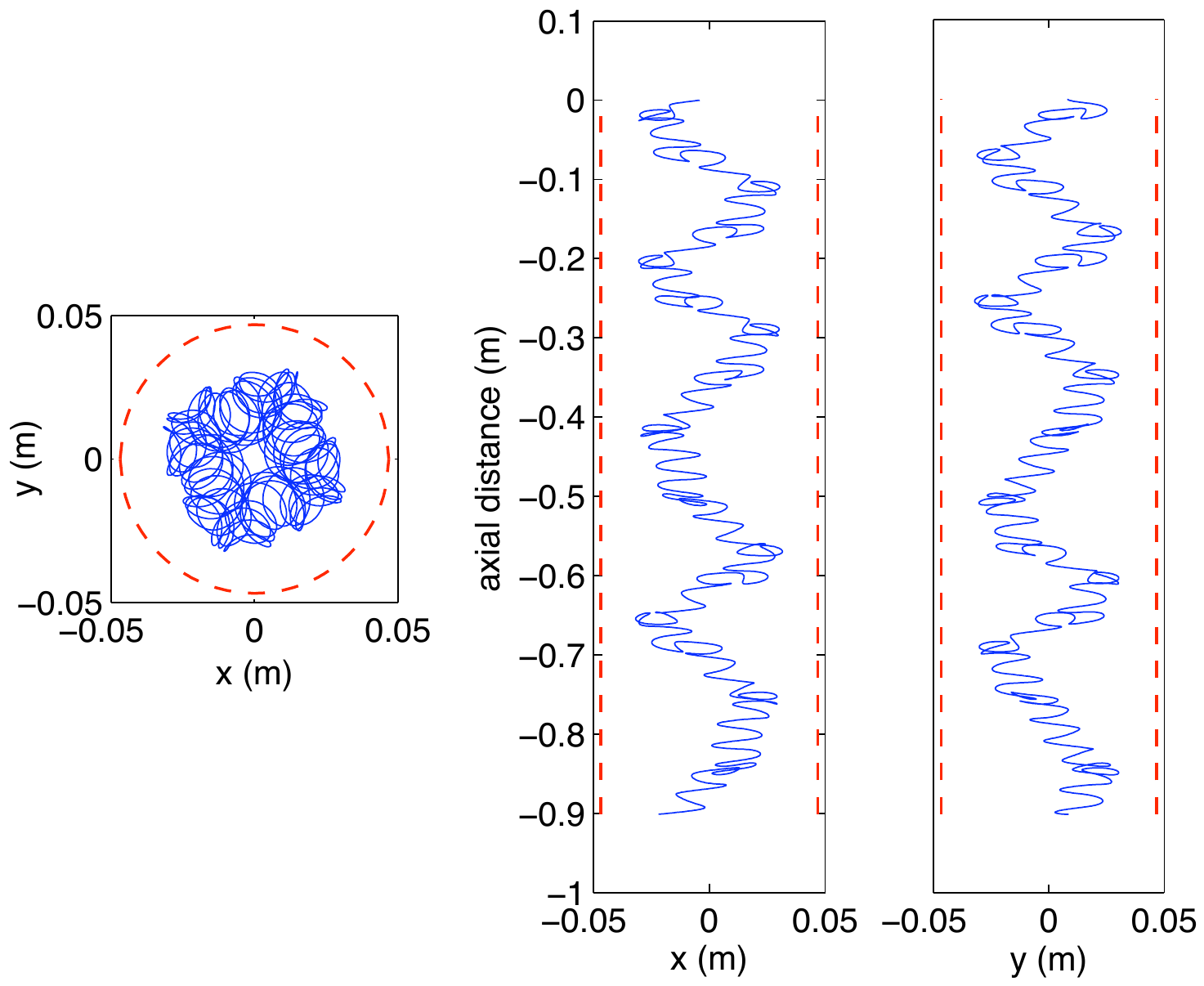}
\end{center}
\caption{Trajectory of a single proton in the magnetic octupole and bias solenoid fields. The dotted lines depict the dimensions of the inner bore of the octupole magnet. }
\label{fig:single_trajectory}
\end{figure}

\begin{figure}[t]
\begin{center}
\subfloat[Magnetic octupole and bias solenoid in an infinitely long configuration.]
{\label{fig:protons_noelectric}\includegraphics[width=0.4\textwidth]{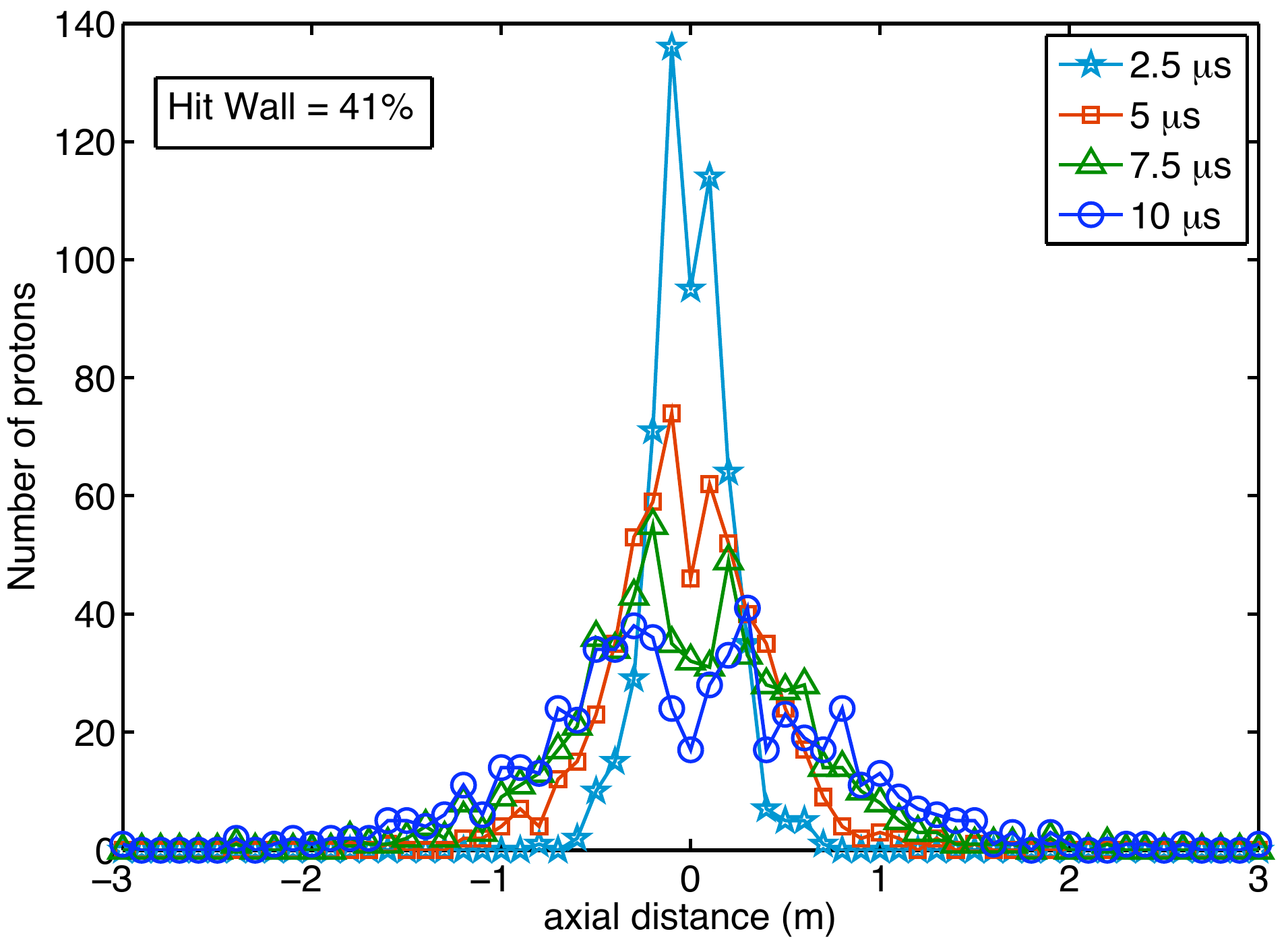}}
\\
\subfloat[Including the electric octupole with alternating electrodes at $\pm$3000$\,$V.]
{\label{fig:protons_3000V}\includegraphics[width=0.4\textwidth]{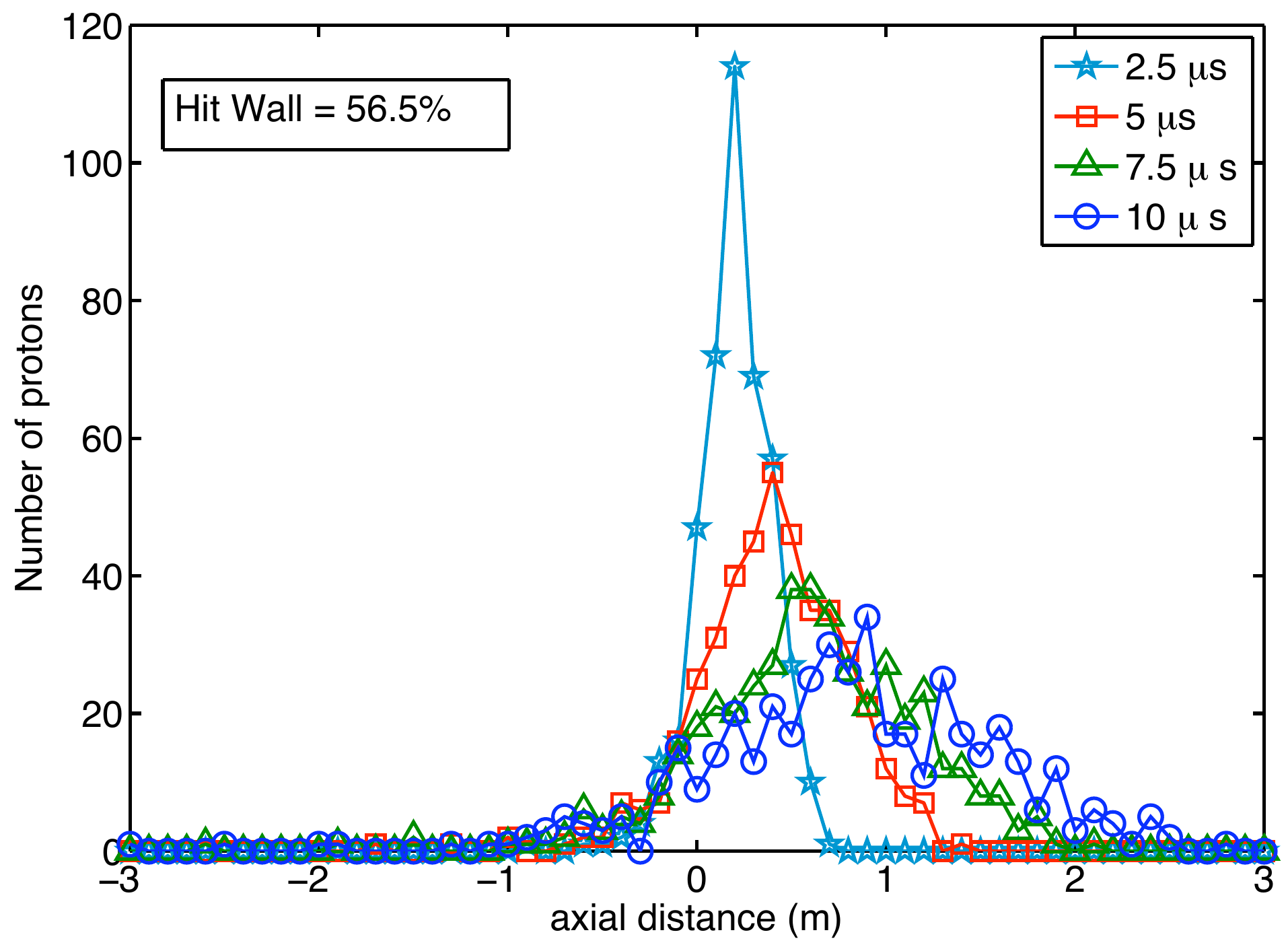}}
\caption{Histogram of the axial distances of 1000 protons at different times during the simulation. The "Hit Wall" parameter is the percentage of protons which have collided with the wall of the octupole after 10$\,\mathrm{\protect \mu s}$ (see text).}
\label{fig:protons_statistical}
\end{center}
\end{figure}

Trajectories of protons were simulated in an infinitely long configuration of octupole and bias fields. A variable step-size Runge-Kutta (4,5) numerical solver was used \cite{Dormand1980a}.  A simulated trajectory demonstrating the behavior described previously is shown in Fig.~\ref{fig:single_trajectory}. Histograms of the axial positions after different times from an ensemble of protons with trajectories starting from $z=0$ are shown in Fig. ~\ref{fig:protons_noelectric}. The results show a large number with a small component of their energy in the axial mode and therefore move only very slowly from their initial $z=0$ positions. To increase the efficiency of the extraction, electrodes are placed in between the magnetic poles (see Fig. \ref{fig:trap3D}) in order to produce electric field lines perpendicular to the magnetic field lines from the octupole. In this way, a $\vec{E}\times \vec{B}$ drift force can be used to direct the protons towards the extraction electrodes. Fig. \ref{fig:protons_3000V} demonstrates the effectiveness of the electric octupole with alternating poles at potentials of $+3$ kV and $-3$ kV.

The "Hit Wall" percentage quoted are the number of protons that collided with the trap wall during the $10$ $\mathrm{\mu s}$ of simulation time. If protons reach the extraction electrodes before this time, which, given the length of the trap, is true for many protons, the loss percentage is actually smaller and the quoted percentages should therefore be considered as pessimistic extraction efficiency estimates.

Around $80$ \% of the wall collisions take place during the first microsecond, due to protons being produced too close and with initial velocities directed towards the trap wall, while the remaining losses proceed at a constant rate of $\sim\,$1$\,-\,$3 \% per microsecond. This disadvantage for the protons is in fact an effective loss mechanism for the decay electrons which can not be extracted due to the large accelerating potentials required. If they remain in the trap for long periods of time, they can produce a time-dependent background of ions. A conservative estimate shows that the mean time for the production of one ion by an electron in a vacuum of $10^{-8} \, \textrm{mbar}$ is $>$ 50$\,\textrm{s}$. The loss rate of protons is much lower than that for electrons due to their higher velocities. Therefore, electrons should relatively quickly be absorbed by the trap walls before they cause a noticeable level of ionization.

\section{Systematics from neutron loss channels}
\label{sec:systematics}

With UCNs magnetically trapped in vacuum, there remain three neutron loss mechanisms aside from $\beta $-decay: neutron depolarization, interaction with residual gas, and marginally trapped neutrons. The probability of depolarization is highest in the low field region in the center of the trap. However, with a minimum field of $0.3\,$T, this probability is reduced to a negligible level. For instance, the rate of depolarization in a quadrupole Ioffe-type trap with a minimum field of $0.2\,$T contributes only a level of $10^{-5}$ to the systematic uncertainty in the extracted $\tau _{\mathrm{n}}$ \cite{Brome2000}. Also, interactions of UCNs with residual gas only affects the storage time to a level of $< 10^{-6}$ if a vacuum of $< 10^{-8}\,$mbar is obtained. Both of these effects can be investigated by monitoring the neutron losses in separate systematic tests.

The most severe channel of neutron loss, and the problem most needed to be addressed in magnetic bottle experiments, is that of marginally trapped neutrons. These are neutrons with energy greater than the trapping potential but exist in quasi-stable orbits and therefore are stored for periods of time comparable to the neutron lifetime. For instance, in Ref. \cite{Ezhov2005,Ezhov2006a} a cleaning time of around 1200$\,$s was required. The required cleaning time increases if the trap geometry is highly symmetrical, if a material reflector is coated on the inner surface of the trap, and if the regions which define the trapping field strength are localised "holes". The latter two factors are reduced in this trap by the use of an absorber on the inner walls of the trap (see Sec. \ref{sec:lifetime_measure}) and by having a larger diameter of the end solenoid (see Sec. \ref{sec:trap_setup}).

A standard procedure of removing above-threshold neutrons in Ioffe-type traps involves ramping down the field of the magnetic trap \cite{Brome2001}. However this induces a large loss of stored neutrons. We can apply this procedure in our setup by ramping the bias field solenoid. Another possible approach is to use the technique of chaotic cleansing \cite{Bowman2005} which involves breaking the symmetry of the trap and accelerating exploration of the whole trap by the neutrons. This can be achieved by activating a current-carrying wire or by inserting a neutron reflector inside the trapping volume for a brief period of time. These techniques can be easily applied in our setup and will be tested in upcoming experiments.

Since the walls of the converter vessel provide an efficient energy cutoff during the long period of UCN accumulation, we can define the maximum energy of the stored UCNs during this stage with an appropriate diameter and wall coating of the vessel. The lower cut-off in the UCN spectrum can also be altered by changing the material of the extraction foil. Systematic changes in the storage time with different neutron energy spectra can therefore be investigated.

\section{Estimated Lifetime Precision}

The new beam H172 at the ILL, after $9\,\mathrm{\mathring{A}}$ monochromation using intercalated graphite crystals, offers a neutron flux of $\mathrm{d}\phi /\mathrm{d}\lambda =3\times 10^{9}\,\mathrm{cm^{-2}s^{-1}nm^{-1}}$. An estimate results in a number of $\sim 4\times 10^{5}$ neutrons stored per filling procedure. This provides a statistical uncertainty in the neutron lifetime of $\sim 3\,$s per fill if valid separate measurements of the background are made. This estimate assumes 30\% efficiency for the extraction through the thin foil and 70\% efficiency for proton detection. After one reactor cycle at the ILL, a statistical uncertainty of $\sim 0.1\,$s should be reached.

\section{Acknowledgements}

We gratefully acknowledge support from the German BMBF (contract number 06MT250) and the ESFRI ILL 20/20 project for funding the current work.

\bibliographystyle{elsart-num}
\bibliography{../bib}

\end{document}